\documentclass[aps,showpacs,amsmath,amssymb,nofootinbib,preprintnumbers]{revtex4}

\usepackage{graphicx}
\usepackage{color}

\def \be  {\begin{equation}}
\def \ee  {\end{equation}}
\def \bea {\begin{eqnarray}}
\def \eea {\end{eqnarray}}

\begin{document}
\preprint{ECTP-2010-07}
\title{Hubble Parameter in QCD Universe for finite Bulk Viscosity}

\author{A.~Tawfik}
\email{drtawfik@mti.edu.eg}
\affiliation{Egyptian Center for Theoretical Physics (ECTP), MTI University,
 Cairo-Egypt}
 \author{M.~Wahba}
\affiliation{Egyptian Center for Theoretical Physics (ECTP), MTI University,
 Cairo-Egypt}
\author{H.~Mansour}
\affiliation{Department of Physics, Cairo University, Giza-Egypt}
\author{T.~Harko}
\affiliation{Department of Physics and Center for Theoretical
and Computational Physics,
The University of Hong Kong, Pok Fu Lam Road, Hong Kong}

\date{\today}

\begin{abstract}
We consider the influence of the perturbative bulk viscosity on the evolution of the Hubble parameter in the QCD era of the early Universe. For the geometry of the Universe we assume the homogeneous and isotropic Friedmann-Lemaitre-Robertson-Walker metric, while the  background matter is assumed to be characterized by barotropic equations of state, obtained from recent lattice QCD simulations, and heavy--ion collisions, respectively. Taking into account a perturbative form for the bulk viscosity coefficient, we obtain the evolution of the Hubble parameter, and we compare it with its evolution for an ideal (non--viscous) cosmological matter. A numerical solution for the viscous QCD plasma in the framework of the causal Israel-Stewart thermodynamics is also obtained. Both the perturbative approach and the numerical solution qualitatively agree in reproducing the viscous corrections to the Hubble parameter, which in the viscous case turns out to be slightly different as compared to the non--viscous case. Our results are strictly limited within a very narrow temperature-- or time--interval in the QCD era, where the quark-gluon plasma is likely dominant.

\end{abstract}



\maketitle

\section{Introduction}\label{sec:intro}

The bulk viscosity is assumed to play an essential role in various eras of the early Universe \cite{tawTD}. Causal bulk viscous thermodynamics has been extensively used for describing the dynamics of the early Universe, and in astrophysical applications. Due to the complicated nature of the cosmological evolution equations in the early Universe, very few exact solutions are known. Isotropic homogeneous Universes filled with causal viscous fluid obeying the relation $\xi \sim \rho^{s}$, where $\xi $ is the bulk viscosity coefficient, and $\rho$ is the energy density, have been studied for special cases of the exponent $s$, $s=1$ and $s=1/2$, respectively, in \cite{earlyQGP} and \cite{ChJa97,MaHa99b,MaTr97}, respectively. Arbitrary values of the exponent $s$ are also considered in Refs. \cite{MaHa99a,earlyQGP}. It has been proposed that causal bulk viscous thermodynamics can give a model for the phenomenological matter creation in the early Universe \cite{earlyQGP,ChJa97}.

Recent RHIC results apparently indicate that hot and dense matter has been formed in heavy-ion collisions~\cite{reff1}, which likely agrees with the {\it new state of matter} predicted in lattice QCD simulations~\cite{reff5,mueller2} and clearly points out that the bulk viscosity $\xi$ is non-negligible in regions close to the QCD critical temperature, $T_c$. Nevertheless, there has been a debate on how to  modify the cosmological standard model \cite{harrison}, and the possibility of applying the QCD barotropic equations of state (EoS) \cite{barotrp}  in the QCD era of the early Universe \cite{earlyQGP,TawCosmos}. Reasons for such a {\it ''categorical resistance''} might be the mathematical difficulties associated with the second Abel type non-homogeneous and non-linear differential equations describing the dynamics of causal bulk viscous models~\cite{earlyQGP,TawCosmos,maartensR}. The long history of cosmological studies in which background matter has been modeled as an ideal (non-viscous) fluid would represent another reason. Obviously, the viscous characteristics of cosmological matter likely leads to important consequences in cosmology and astronomy~\cite{taw08}.

It is the purpose of the present work to investigate the QCD cosmological era of the evolution of the Universe. We assume that the Universe, described by the standard  Friedmann-Lemaitre-Robertson-Walker (FLRW) cosmic background geometry, is filled with a relativistic viscous QCD plasma, whose bulk viscosity is supposed to be finite. The equation of state of the plasma  is obtained from recent heavy-ion collisions and lattice QCD simulations, respectively ~\cite{karsch07,Cheng:2007jq}. The QCD plasma is assumed to be formed at temperatures in the range $0.2\leq T\leq10\;$GeV, or in the time--period $18.35\leq t \leq 0.0073\;$GeV$^{-1}$, respectively.

To fully describe the effects of bulk viscosity on the cosmological evolution we analyze several scenarios. Firstly, we assume a perturbative bulk viscosity, $\xi=\alpha$, where $\alpha$ is a constant independent of $T$ and/or $\rho$. However, this assumption  has  an insufficient thermodynamical motivation. Nevertheless, in the absence of a better alternative, we can consider this approach in order to obtain some indications for the deviation from the zero bulk viscosity case $\xi=0$. On the other hand the assumption of the constant bulk viscosity coefficient is physically motivated,  if perturbative methods are used to describe bulk viscous processes. Thus, the $T$-- and/or the $\rho$--dependence of $\xi$ or $\alpha$ can be ignored.  A better approach could be the use of ${\cal N}=4$ SYM \cite{baier}, which is apparently valid at very high $T$.

The main concern about considering a constant $\xi$ stems from the corresponding $T$-- and/or $\rho$--dependence of the relaxation time $\tau$, and particle decay width $\Gamma$, which has to ensure that the speed of viscous pulses does not exceed the speed of light. For example, a $T$--dependence of $\tau$ of the form  $\tau=(2-\ln2)/(2\pi T)$, guarantees a positive heat capacity. A constant relaxation time can be interpreted as indicating  that the relaxation towards the thermal equilibrium is very slow. On the other hand, both the $1/T$ and the $\tau$ times scale become arbitrarily small with decreasing time $t$. At $T>T_c$, the quark masses are much smaller than $T$. Then the only way to construct a characteristic time scale for $t$ is $1/T$. Also, powers of dimensionless coupling $\alpha_s$ can be multiplied by $1/T$. Once again, the perturbative methods are consistent with such an approximation.

Secondly, we consider numerical solutions for the evolution equation in the case of  finite viscous background matter characterized by the Israel-Stewart causal thermodynamics. Barotropic EoS for both $\xi$ and $\tau$ are adopted from recent lattice QCD simulations and quasi--particle (QP) effective models \cite{qpm}, respectively. The numerical methods are very much sensitive to the boundary conditions. To keep this sensitivity as small as possible, we use the results obtained from the perturbative treatment.

The present paper is organized as follows. The QCD plasma model is described in Section~\ref{sec:model}. The perturbative treatment and the numerical solution are considered in Section~\ref{sec:new}. We discuss and conclude our results on the evolution of the Hubble parameter in the QCD era in Section \ref{sec:concl}.

\section{Geometry, field equations, and consequences}
\label{sec:model}

In spherical coordinates $\left(t,r,\theta , \phi \right)$, the line element of a homogeneous and isotropic flat Universe can be represented in the standard FLRW form as
\begin{equation}  \label{1}
ds^{2}=dt^{2}-a^{2}(t) \left[dr^{2}+r^{2}\left( d\theta
^{2}+\sin ^{2}\theta d\phi^{2}\right) \right],
\end{equation}
where $a$ is the scale factor. The Hubble parameter is defined as $H=\dot a/a$.
For a vanishing cosmological constant $\Lambda $, the dynamics of the Universe is described by the Einstein gravitational field equations,  given by
\begin{equation}  \label{ein}
R_{ik}-\frac{1}{2}g_{ik}R=8\pi\; G\; T_{ik}.
\end{equation}
The energy-momentum tensor of the bulk viscous cosmological fluid filling the
very early Universe is given by
\begin{equation}
T_{i}^{k}=\left(  \rho +p+\Pi\right)  u_{i}u^{k}-\left(  p+\Pi\right)
\delta_{i}^{k},\label{1_a}%
\end{equation}
where $i,k$ takes the values $0,1,2,3$, $\rho$ is the mass density, $p$ the thermodynamic pressure, $\Pi $ the
bulk viscous pressure, and $u_{i}$ is the four velocity, satisfying the condition
$u_{i}u^{i}=1$. The particle and entropy fluxes are defined according to
$N^{i}=nu^{i}$ and $S^{i}=sN^{i}-\left(  \tau\Pi^{2}/2\xi T\right)
u^{i}$, where $n$ is the number density, $s$ the specific entropy, $T\geq0$ the
temperature, $\xi$ the bulk viscosity coefficient, and $\tau\geq0$ the
relaxation coefficient for transient bulk viscous effect (i.e. the relaxation time), respectively.
In the following we shall also suppose that the energy-momentum tensor of the
cosmological fluid is conserved, that is $T_{i;k}^{k}=0$.

The bulk viscous effects can be generally described by means of an effective
pressure $\Pi $, formally included in the effective thermodynamic pressure $%
p_{eff}=p+\Pi $ \cite{maartensR}. Then in the comoving frame the energy momentum tensor
has the components $T_{0}^{0}=\rho ,T_{1}^{1}=T_{2}^{2}=T_{3}^{3}=-p_{eff}$.
For the line element given by Eq.~(\ref{1}), the Einstein field equations read
\begin{eqnarray}  \label{2n}
\left( \frac{\dot{a}}{a}\right)^{2} &=& \frac{8\pi}{3}G \;\rho, \\
\frac{\ddot{a}}{a} &=& -\frac{4\pi}{3}G \; \left( 3p_{eff}+\rho \right),
\label{3}
\end{eqnarray}
where the dot denotes the derivative with respect to the time $t$, and $G$ is the gravitational constant. The energy density of the cosmic matter fulfills the conservation law:
\begin{equation}  \label{5n}
\dot{\rho}+3H\left( p_{eff}+\rho \right) =0.
\end{equation}

In presence of bulk viscous stress $\Pi $, the effective
thermodynamic pressure term becomes $p_{eff}=p+\Pi $. Then Eq.~(\ref{5n}) can be written as
\begin{equation}  \label{6n}
\dot{\rho}+3H\left( p+\rho \right) =-3\Pi H.
\end{equation}

For the evolution of the bulk viscous pressure we adopt the causal evolution
equation \cite{maartensR}, obtained in the simplest way (linear in $\Pi)$ to
satisfy the $H$-theorem (i.e., for the entropy production to be non-negative,
$S_{;i}^{i}=\Pi^{2}/\xi T\geq0$ \cite{Is76,IsSt76}). According to the causal relativistic
Israel-Stewart theory, the evolution equation of the bulk viscous pressure reads~\cite{maartensR}
\begin{equation}  \label{8n}
\tau \dot{\Pi}+\Pi =-3\xi H-\frac{1}{2}\tau \Pi \left( 3H+\frac{\dot{\tau}}{%
\tau }-\frac{\dot{\xi}}{\xi }-\frac{\dot{T}}{T}\right).
\end{equation}
In order to  close the system of equations (\ref{2n}), (\ref{6n}) and (\ref{8n})
we have to add the equations of state for $p$ and $T$. The equation of state, the temperature and the bulk
viscosity of the quark-gluon plasma (QGP), can be determined approximately at high temperatures from recent lattice QCD calculations as~\cite{Cheng:2007jq,karsch07}
\begin{equation}\label{13an}
P = \omega \rho,\hspace*{1cm}T = \beta \rho^r,\hspace*{1cm}\xi = \alpha \rho + \frac{9}{\omega_0} T_c^4,
\end{equation}
\begin{equation}\label{13bn}
\alpha = \frac{1}{9\omega_0}  \frac{9\gamma^2-24\gamma+16}{\gamma-1},
\end{equation}
where $\omega = (\gamma-1)$, $\gamma \simeq 1.183$, $r\simeq 0.213$, $\beta\simeq 0.718$, and  $\omega_0 \simeq 0.5-1.5$ GeV, respectively. In the following we assume that $\alpha \rho >> 9/\omega_0 T_c^4$, and therefore we take $\xi \simeq \alpha \rho$. In order to close the system of the cosmological equations, we have also to give the expression of the relaxation time $\tau $, for which we adopt the expression \cite{maartensR},
\begin{equation}\label{tau}
\tau=\xi\rho^{-1}\simeq\alpha .
\end{equation}

Eqs.~(\ref{13an})
are standard in the study of the viscous cosmological models, whereas the equation for $\tau$ is a
simple procedure to ensure that the speed of viscous pulses does not exceed
the speed of light. Eq. (\ref{tau}) implies that the relaxation time in our treatment is constant, but strongly depends on the EoS. These equations are without sufficient thermodynamical motivation,
but in the absence of better alternatives, we shall follow the practice of adopting
them in the hope that they will at least provide some indication of the range of
bulk viscous effects. The temperature law is the simplest law guaranteeing positive
heat capacity.

\section{Perturbative and numerical solutions of the field equations}\label{sec:new}

For a viscous FLRW Universe filled with a viscous quark--gluon plasma (QGP), the evolution equation of the Hubble parameter is given by
\begin{eqnarray}\label{eq:1}
\alpha H\ddot H+\frac{3}{2}[1+(1-r)\gamma]\alpha H^2\dot H+H\dot H-(1+r)\alpha\dot H^2+\frac{9}{4}(\gamma-2)\alpha H^4+\frac{3}{2}\gamma H^3 = 0.
\end{eqnarray}
The parameters, $\alpha$, $r$ and $\gamma$ are the coefficients and the exponents of the barotropic QGP EoS, discussed later in Eqs. (\ref{13an})--(\ref{13bn}) \cite{earlyQGP}. In the limit of a vanishing $\alpha$, a non--viscous evolution equation can be obtained from Eq.~(\ref{eq:1}). Then,
\begin{eqnarray}\label{eq:2}
H_0(t)\dot H_0(t)+\frac{3}{2}\gamma H^3_0(t)=0,
\end{eqnarray}
where $H_0(t)$ is the Hubble parameter corresponding to the non--viscous background matter. Eq.~(\ref{eq:2}) can easily be solved by
\begin{equation}\label{eq:3}
H_0(t)= \frac{2}{3\gamma}\,t^{-1},
\end{equation}
The mathematical solution $H_0=0$ of Eq.~(\ref{eq:2}), which can be interpreted physically as describing a static Universe with a vacuum background geometry, has been ignored, as irrelevant to the present case. The second solution refers to a dynamic state. The acceleration (deceleration) of the Universe reads
\begin{eqnarray}
\dot H_0(t) &=& \frac{\ddot a}{a}-H_0^2(t),\\
\ddot a &=& a\left(-\frac{2}{3\gamma}\frac{1}{t^2}+H_0^2(t)\right)\label{eq:4},
\end{eqnarray}
implying that the type of  the expansion of the Universe is exclusively determined by $\gamma$. For $\gamma>2/3$, the Universe decelerates. As shown below, in a quark-gluon plasma $\gamma$ could be as large as two times this threshold. For the viscous QGP model, the expressions for $H$ will be obtained by means of a perturbative analysis and by considering numerical methods.

\subsection{Perturbative Bulk Viscosity}

In the following we consider the effects of the inclusion of the bulk viscosity in the cosmological standard model \cite{harrison} by means of a perturbative approach. It is essential to highlight that the validity of this treatment is restricted to the QCD era. It is considered that the QCD era lasted a much shorter time interval than the era corresponding to a finite scale and a strong running coupling $\alpha_s$ \cite{alfasA,alfasB}. As a first step in our analysis we obtain perturbatively the deviations of the Hubble parameter from the non--viscous cosmological picture.

\subsubsection{A First--Order  Perturbative Correction}

By using a first--order perturbative correction, $f(t)$, the solution of Eq.~(\ref{eq:1}) can be written as
\begin{equation}\label{eq:6}
H(t)=H_0(t)+\alpha f(t),
\end{equation}
where $H_0(t)$ is the Hubble parameter corresponding to a vanishing bulk viscosity. Substituting this into Eq. ~(\ref{eq:1}) leads to
\begin{eqnarray}\label{eq:7}
&& \alpha H_0(t)\ddot H_0(t)+\frac{3}{2}[1+(1-r)\gamma]\alpha H^2_0(t)\dot H_0(t)+H_0(t)\dot H_0(t)+\alpha H_0(t)\dot f(t)\nonumber\\
&& +\alpha f(t)\dot H_0(t)-(1+r)\alpha\dot H^2_0(t)+\frac{9}{4}(\gamma-2)\alpha H^4_0(t)+\frac{3}{2}\gamma H^3_0(t)+\frac{9}{2}\alpha\gamma H^2_0(t)f(t)=0.
\end{eqnarray}
This can be written in the form
\bea \label{eq:8}
\dot f(t)+G_1(t)f(t)+G_0(t)&=&0,
\eea
where
\begin{eqnarray}
G_0(t) &=& -\ddot H_0(t)-\frac{3}{2}[1+(1-r)\gamma]H_0(t)\dot H_0(t)-\frac{\dot H_0(t)}{\alpha}+(1+r)\frac{\dot H^2_0(t)}{H_0(t)}
   -\frac{9}{4}(\gamma-2)H^3_0(t)-\frac{3}{2}\gamma\frac{H^2_0(t)}{\alpha}, \\
G_1(t) &=& \frac{\dot H_0(t)}{H_0(t)}+\frac{9}{2}\gamma H_0(t).
\end{eqnarray}
To eliminate $H_0(t)$ and its derivatives, we can use Eqs.~(\ref{eq:3}), (\ref{eq:4}) and
\bea
\ddot H_0(t) &=& \frac{\dddot a}{a} -3H_0(t) \frac{\ddot a}{a} + 2H^3_0(t) = \frac{4}{3\gamma}\frac{1}{t^3},
\eea
respectively, thus closing the coupled set of equations needed to determine the non--homogeneous coefficients $G_0(t)$ and $G_1(t)$. Then
\begin{eqnarray}
G_0(t) &=& A\,t^{-3},\label{eq:9}\\
G_1(t) &=& 2\,t^{-1}\label{eq:10},
\end{eqnarray}
where the coefficient $A$ is given by
\begin{equation}
A=\frac{2}{3\gamma^2}[1+(1-r)\gamma]-\frac{4}{3\gamma}-\frac{2}{3\gamma}(1+r)
-\frac{2}{3}\frac{\gamma-2}{\gamma^3}.
\end{equation}

Equation~(\ref{eq:8}) is a first--order linear differential equation and has the general solution given by
\begin{equation}\label{eq:11}
f(t)= e^{-F}\left(\int e^F G_0(t) dt+C\right),
\end{equation}
where $F=\int G_1(t) dt$, and $C$ is an arbitrary integration constant. Substituting Eqs.~(\ref{eq:9}) and (\ref{eq:10}) into Eq.~(\ref{eq:11}) leads to,
\begin{equation}\label{eq:1ft}
f(t)=t^{-2}[A \ln(t)-C].
\end{equation}
Therefore, the solution of Eq.~(\ref{eq:1}) reads
\begin{equation}\label{eq:12}
H(t)=\frac{2}{3\gamma}\frac{1}{t}+\alpha t^{-2}[A \ln(t) - C].
\end{equation}
Since $C=A\,\ln(t_0)$ does not depend on $t$, the acceleration/deceleration is obtained as
\bea\label{eq:hdot1}
\dot{H}(t) &=& -\frac{2}{3\gamma t^2} + \frac{\alpha}{t^3}\left[ A - 2\left(A\,\ln(t)-C\right)\right].
\eea

\subsubsection{A Second--Order Perturbative Correction}

\begin{figure}[htb]
\includegraphics[angle=0,width=8.5cm]{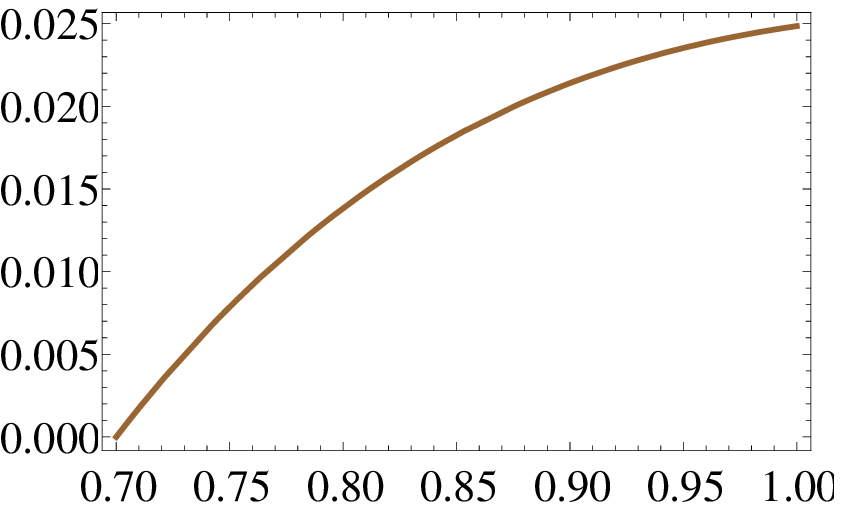}
\includegraphics[angle=0,width=8.65cm]{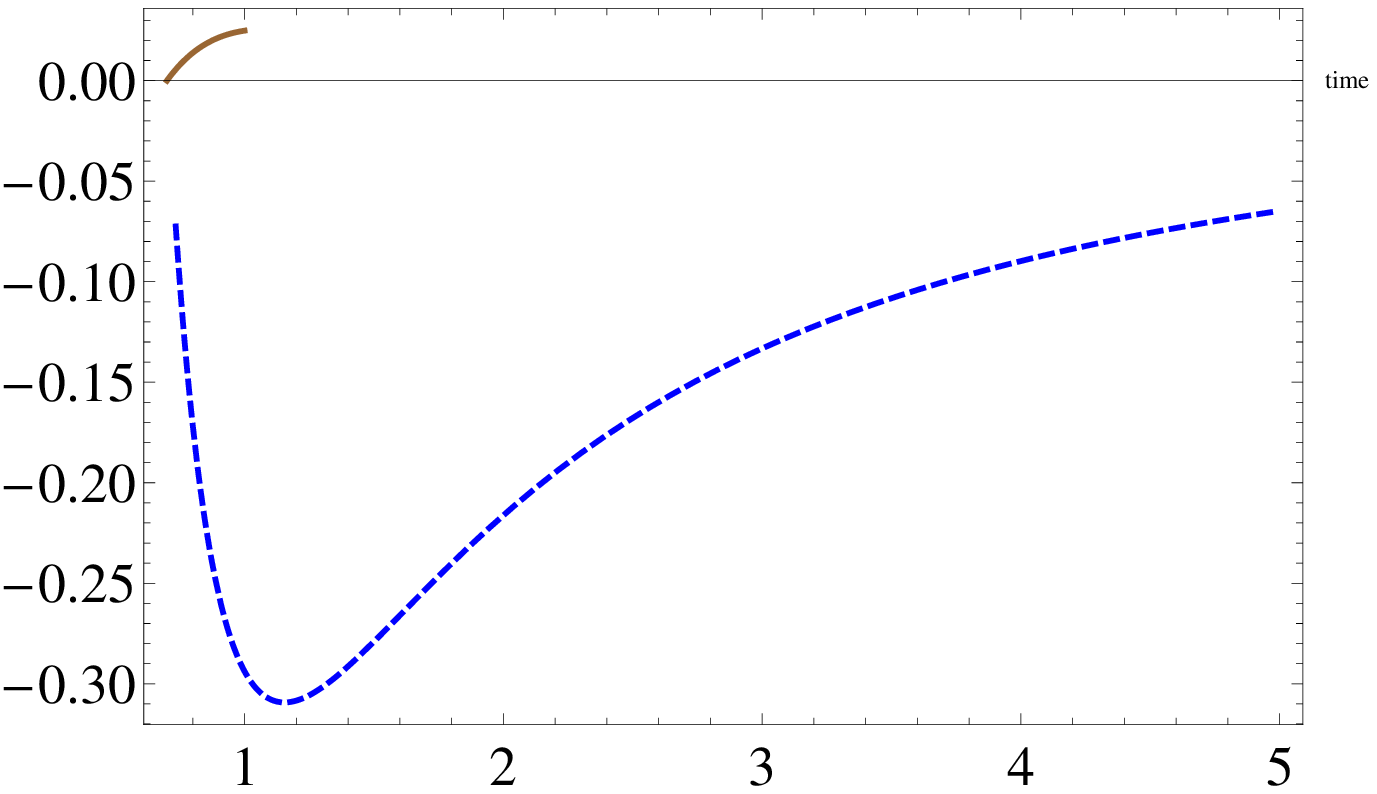}
\caption{\normalsize Left panel: second particular solution $g(t)$ as a function of time $t$.
Only its real part is taken into account, which obviously exists only in $t\in[t_0,1]\,$GeV$^{-1}$. Right panel shows a comparison between first (dashed curve) and second (solid curve) particular solutions. }
\label{fig:1a2}
\end{figure}

A second--order perturbative correction to the Hubble parameter can be represented as
\begin{equation} \label{eq:13}
H(t)=H_0(t)+\alpha f(t)+\alpha^2 h(t).
\end{equation}
In order to solve the resulting evolution equation for the new perturbation function $h(t)$, we can follow the same procedure used in the previous Section. Substituting Eq.~(\ref{eq:13}) and its derivatives into Eq.~(\ref{eq:1}) reduces the problem to a second--order non--homogeneous differential equation given by
\bea\label{eq:2nonh}
f_2(t)\,\ddot h(t)+f_1(t)\,\dot h(t)+f_0(t)\,h(t)&=&k(t),
\eea
where the coefficients of the non--homogeneous equations are
\bea
f_2(t) &=& \frac{1}{3} \alpha^4\,A\, \frac{\ln(t)}{t}, \\
f_1(t) &=& \frac{2}{t}(1+r) f_2(t) + \frac{2\alpha^3}{3\gamma^2t^2}\left[1+\gamma(1-r)\right],\\
f_0(t) &=& \frac{4\alpha^2}{3t^3\gamma^3}\left[\gamma^2 t+\alpha(-4+\gamma(2+\gamma))\right] - \frac{1}{t^2}\left(f_1(t)-\frac{2}{t}(1+r)f_2(t)\right)\left[2t-3\gamma A+6\gamma C + 6A\ln(t)\right],\\
k(t) &=& -\alpha\,f_2(t)- \nonumber \\
 && \frac{f_2(t)}{\alpha^2\gamma^3 t^4} \left[-16+\gamma(8+\gamma(8-8r+3\gamma A+6\gamma C))\right]-\nonumber\\
 && \frac{3f_2(t)}{\alpha\gamma^2 t^4} \left[\gamma^2 A(4r-1)+4C(\gamma(3+\gamma-2r\gamma)-6)\right]-\nonumber \\
 && \frac{3\,A\,\ln(t)\,f_2(t)}{\alpha^2\gamma^2 t^4}  \left[\gamma^2 t + 2 \alpha(-6+\gamma(3+\gamma-2r\gamma))\right]- \nonumber\\
 && \frac{\alpha^2}{3\gamma^3 t^5} \left[8C(\gamma-r-1)+2\gamma^2 A(2r-3)+6\gamma^3C(A+C)\right]- \\
 && \frac{\alpha^2}{3\gamma^3 t^3} \left[-8+2\gamma(2+\gamma(2-2r+3\gamma A))\right]-\nonumber\\
 &&\frac{3\alpha^4}{\gamma t^5} \left[\gamma^2 A(4rC-C-rA-A)+2C^2(\gamma(3+\gamma-2r\gamma)-6) \right]+\nonumber \\
 && \frac{4(1+\gamma-r\gamma)}{3\alpha^3\gamma^3t^3 A \ln(t)}\,f_2(t) + \frac{2(1+\gamma-r\gamma)}{3\alpha^2\gamma^2t^4 A \ln(t)}\,f_2(t) + \frac{3(1+\gamma-r\gamma) C}{\alpha^2\gamma t^5 A \ln(t)}\,(-2A+c(4+\alpha))\,f_2(t) + \nonumber \\
 && \frac{1+\gamma-r\gamma}{\alpha^2\gamma^2t^5}\,f_2(t)\left[8t+6\gamma(-A+ C(4+\alpha))+3\gamma A (4+\alpha) \ln(t)\right] \nonumber
\eea
We assume that $f(t)$, given by Eq.~(\ref{eq:1ft}), represents a particular solution of the following second--order homogeneous differential equation:
\bea\label{eq:2h}
f_2(t)\,\ddot g(t)+f_1(t)\,\dot g(t)+f_0(t)\,g(t)&=&0.
\eea
The solution $g(t)$ of Eq.~(\ref{eq:2h}) reads
\bea
g(t) &=& f(t)\int_{t_0}^t \frac{\exp(-F)}{f^2(t)}\,dt,
\eea
where $F=\int_{t_0}^t f_2(t)/f_1(t)dt$. Then
\bea
g(t) 
     &=& \frac{A \ln(t)-C}{t^2}\,\int_{t_0}^t \frac{t^{4-\beta_2/\beta_1}\left(\ln(t)\right)^{-\beta_3/\beta_1}}{[A \ln(t)-C]^2}\,dt,\\
     \eea
where we have denoted
\bea
\beta_1=\frac{1}{3}\alpha^4 A, \hspace*{.5cm} && \beta_2=\frac{2}{3}\alpha^4 A (1+r),  \hspace*{1.25cm} \beta_3=\frac{2}{3}\frac{\alpha^3}{r^2} (1+r-r\gamma).\nonumber
\eea
A second particular solution for the homogeneous differential equation is given by
\bea \label{eq_2ps}
g(t) &=& \frac{A \ln(t)-C}{t^2}\,\left\{-(2r-3)^{-(1+B)}\; \Gamma\left[1+B,(2r-3)\ln(t)\right]  \right\},
\eea
where $\Gamma$ is the incomplete gamma function, and
\bea
B&=& 2\frac{r(\gamma -1)-1}{\alpha r^2 A}.
\eea
The numerical values of this second--order perturbative correction are negligibly small. Fig.~\ref{fig:1a2} shows the real part of Eqs.~(\ref{eq_2ps}). Obviously, it is valid up to $t=1$.

The general solution of the non--homogeneous differential equation, Eq.~(\ref{eq:2nonh}), can be obtained with the use of the Wronskian determinant, $W=\exp(-F)$,
\bea
h(t) &=& c_1f(t)+c_2g(t) + g(t)\int f(t)\frac{k(t)}{f_2(t)}\frac{dt}{W} - f(t)\int g(t)\frac{k(t)}{f_2(t)}\frac{dt}{W},
\eea
where $c_1$ and $c_2$ are arbitrary constants of integration.

Hence, we conclude that the first--order perturbative correction $h(t)$ seems to give the dominant term, at least according to the quantitative comparison with the numerical solution given in the next Section. The function  $g(t)$ likely makes a very little contribution to the final results. Taking the non--viscous case as a particular solution leads to a similar conclusion. Therefore, there is no need to include the second order correction in Fig.~ \ref{fig:1a1}.

\subsection{Numerical solution for the causal cosmological quark-gluon plasma}

\begin{figure}[htb]
\includegraphics[angle=-90,width=8.5cm]{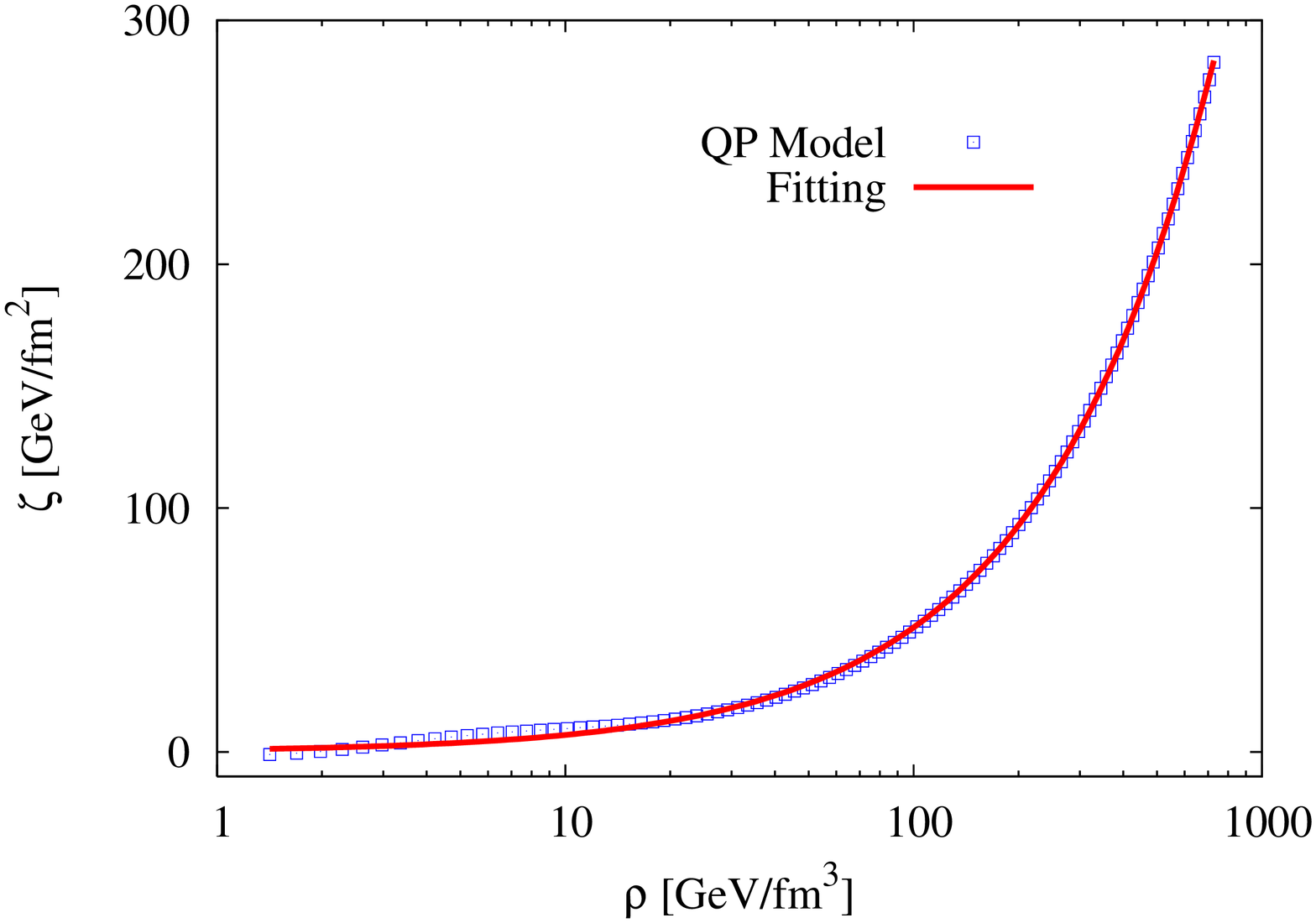}
\includegraphics[angle=-90,width=8.5cm]{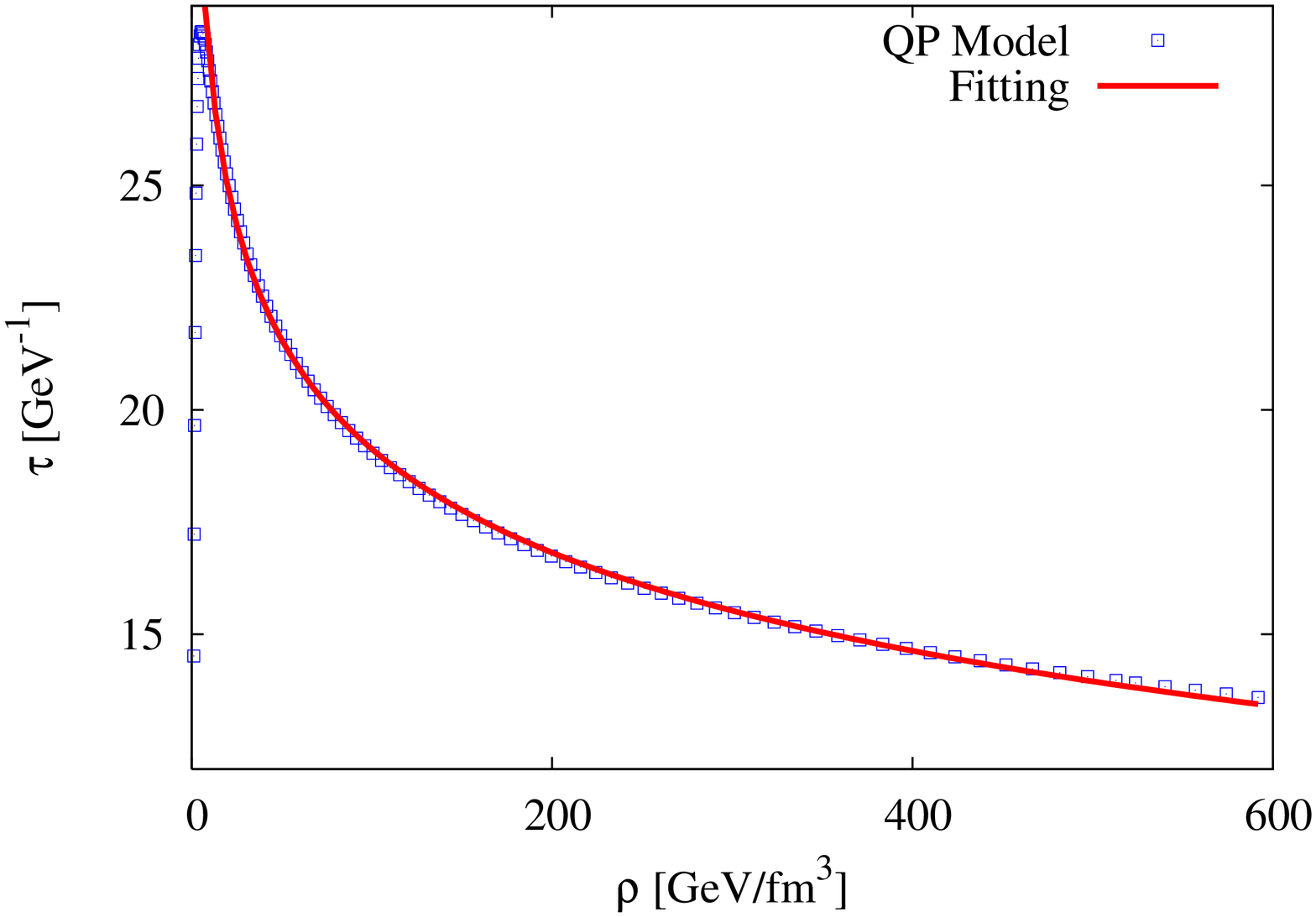}
\caption{\normalsize The dependence of bulk viscosity $\xi$ on energy density $\rho$ in QGP matter for the quasi--particle model. The right panel shows the $\rho$--dependence of the relaxation time $\tau$ in GeV$^{-1}$ from the same model.}
\label{fig:tau}
\end{figure}

The inclusion of the bulk viscous effects can generally be done through an effective
pressure $\Pi$, which is formally included in the effective thermodynamic pressure $p_{eff}$, where $p_{eff}=p+\Pi $, and $p$ is the thermodynamic pressure.
EoS's for $p$ and $T$ are necessary in order to close the system of equations Eq.~(\ref{2n}), (\ref{6n}) and (\ref{8n}). $\tau$ and $\xi$ can be  determined by using some  phenomenological approaches. For instance, $\xi$ in QGP matter at high $T$ can be estimated from recent lattice QCD simulations~\cite{karsch07,Cheng:2007jq}. The relaxation time can also be derived from the quasi--particle (QP) model, which is effectively used to reproduce the lattice QCD thermodynamics \cite{qpm}.
\bea
p&=&\omega \rho, \label{13a}\\
T&=&\beta\rho^r, \label{13b}
\eea
where  $\omega=(\gamma-1)$, $\omega_0 \simeq 0.5-1.5$ $GeV$, $r=0.39$ and $\beta=0.718$.
According to recent lattice QCD simulations \cite{karsch07}, it is found that $s=1$.  In the quasi--particle model, it has been found that the bulk viscosity $\xi$ and the relaxation time $\tau$ have the following barotropic forms,
\bea
\xi &=& (0.959\pm 0.006)\; \rho^{(0.863\pm 0.001)}, \\
\tau &=& (64.189\pm 1.667)\; \rho^{(-0.075\pm 0.003)} -(26.276\pm 1.774),
\eea
which are graphically shown in Fig.~\ref{fig:tau}. In the left panel, the dependence of the bulk viscosity (in units of GeV/fm$^{2}$)  on $\rho$ (in units of GeV/fm$^{3}$) is presented. The right panel shows the decay of the relaxation time $\tau$ in QGP matter with increasing $\rho$. The parameters of the quasi--particle model are first adjusted to reproduce the thermodynamics of lattice QCD simulations, and then they are used to calculate both $\xi$ and $\tau$. The thermodynamic consistency of the barotropic relations of $\xi$ and $\tau$ is guaranteed \cite{tawuro}. This can be partly seen from the fact that the ratio $\xi/\tau$ strongly depends on $\rho$.

Substituting Eq.~(\ref{8n}) into Eq.~(\ref{6n}) leads to Eq. (\ref{eq:1}).
When assuming that the last two terms in the right hand side of this equation are vanishing, then we are left with a Bernoulli type differential equation, with the solutions $H=0$ and $H\approx -2\gamma/[3\alpha(2-\gamma)]$. The case $H=0$ corresponds to a vacuum Universe, and hence this solution can be ignored. If it would be possible to express $\xi$ as a function of $t$, we would obtain $2\dot H+3\gamma H^2-3\xi(t)H=0$,
which is a Bernoulli type equation, and has the solution 
\begin{equation}
H(t)=\frac{\exp{\left[\left(3/2\right)\int\xi(t)dt\right]}}{C+\exp{\left[\left(3/2\right)\gamma \int\xi(t)dt\right]}},
\end{equation}
where $C$ is an arbitrary constant of integration.
A numerical solution of Eq.~(\ref{eq:1}) for $H$ is presented in Fig.~\ref{fig:3a1}. It strongly depends on the initial conditions for $H$, $\dot H$ and $C$. As mentioned previously, the initial time is defined as $t_0=0.734\;$GeV$^{-1}$, at which $T=1.0\;$GeV.

\section{Discussion and final remarks}
\label{sec:concl}

\begin{figure}[htb]
\includegraphics[angle=-90,width=8.5cm]{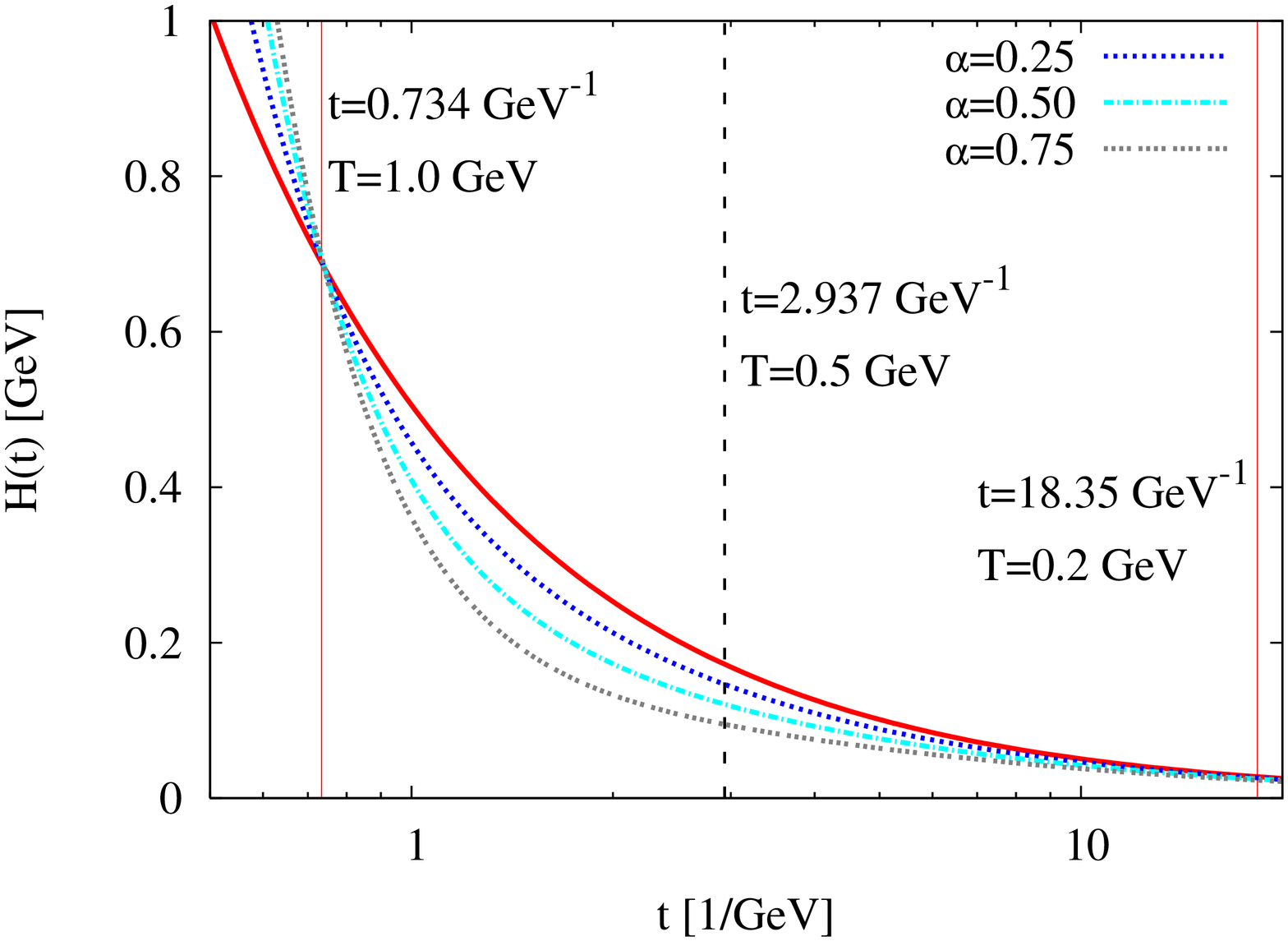}
\includegraphics[angle=-90,width=8.5cm]{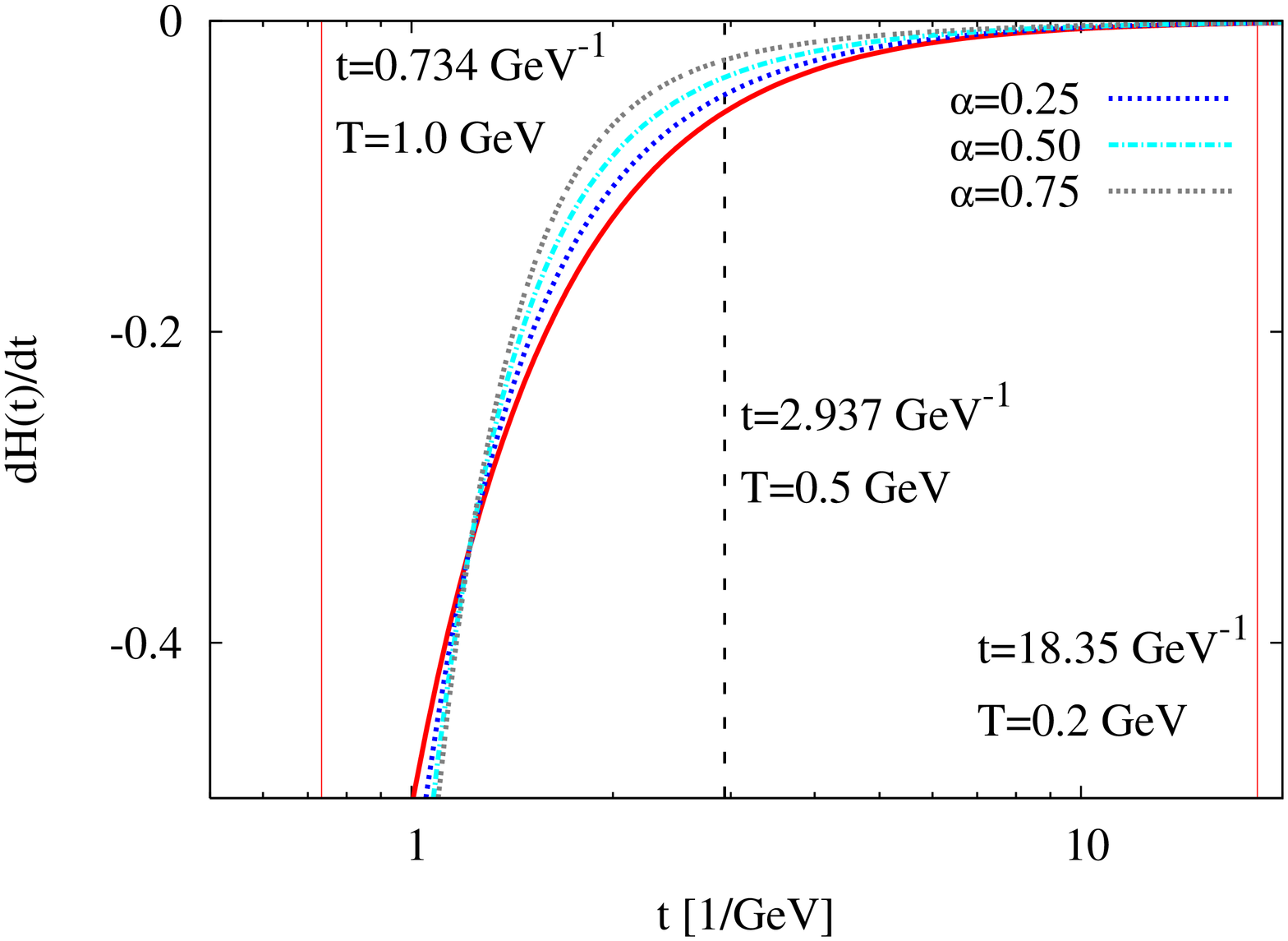}
\caption{\normalsize Left panel: the time evolution of the Hubble parameter $H(t)$ is depicted for
 different values of the first--order perturbative bulk viscosity, $\alpha$. The vertical lines determine the time $t$ and temperature $T$ boundaries of the QCD era in early Universe that have been suggested in present work. In the right panel, $\dot{H}(t)$ is given for different values of $\alpha$. }
\label{fig:1a1}
\end{figure}

According to the standard cosmological standard the QCD era is characterized by a temperature range of-- $0.2<T<10\;$GeV, or a time--interval of $18.35\leq t \leq 0.0073\;$GeV$^{-1}$. In Fig.~\ref{fig:1a1}, the  boundaries of the QCD era are drawn by vertical lines. The dashed one refers to $T=0.5\;$GeV, or $t=2.937\;$GeV$^{-1}$, respectively. The right and the left vertical lines give the boundaries, $T=0.2\;$GeV or $t=18.35\;$GeV$^{-1}$ and $T=1.0\;$GeV or $t=0.734\;$GeV$^{-1}$, respectively.

\begin{figure}[htb]
\includegraphics[angle=0,width=7.cm]{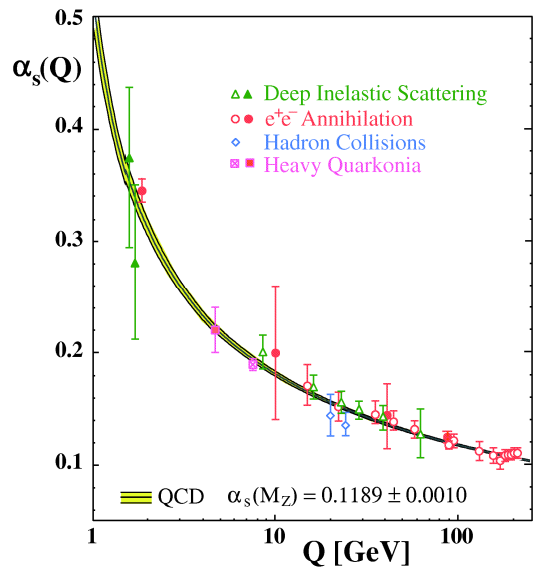}
\caption{\normalsize Various measurements of strong running coupling $\alpha_s(Q)$ as a function of the energy scale
$Q$ in GeV units. The graph is taken from Ref. \cite{alfasA}. The curves represent the QCD predictions, and their systematic certainty.}
\label{fig:alfas}
\end{figure}

In principle, the QCD era is defined according to the energy scale available in the early Universe, and according to the asymptotic behavior of the partonic matter. Although QCD does not predict the absolute size of the strong running coupling $\alpha_s$, it makes a precise prediction for its energy--dependence. Since $\alpha_s$ is the most important degree of freedom of QCD, we define the region for which our treatment is consistent as the region where $\alpha_s$ remains finite. It is important to mention that this degree of freedom appears even if the quarks are entirely excluded. The scale--dependence of the coupling constant $\alpha_s$ is governed by the function $\beta$, which encodes the running of the coupling, and whose perturbative expansion in the four--loop approximation is \cite{betaalpha}.
\bea
\beta(\alpha_s(Q^2)) &=& - \beta_0\,\alpha_s^2(Q^2) - \beta_1\,\alpha_s^3(Q^2) - \beta_2\,\alpha_s^4(Q^2) - \beta_3\,\alpha_s^5(Q^2) + \cdots,
\eea
where the coefficients $\beta_0$, $\beta_1$, $\beta_3$, and $\beta_3$ have been determined by  perturbative methods, and found to be dependent on the number of active quark flavors $n_f$ at the corresponding energy scale $Q$. For simplicity, let us consider the one--loop approximation. Then
\bea\label{eq:alfaQCD}
\alpha_s(Q^2) &\simeq& - \frac{12\pi}{(33-2n_f)\;\ln\left(Q^2/\Lambda^2\right)},
\eea
where $\Lambda$ is the QCD energy scale, depending on $n_f$, and on the renormalization scheme. If $Q=\Lambda$, then $\alpha_s$ diverges. On the other hand, $\alpha_s \rightarrow 0$ at very large energy scale. A summary of $\alpha_s$--measurements is graphically displayed in Fig.~\ref{fig:alfas}. It is obvious that the measurements
unambiguously confirm the QCD predictions for $\alpha_s$, i.e, Eq.~(\ref{eq:alfaQCD}), as well as the approach towards asymptotic freedom. The figure is taken from Ref. \cite{alfasA}. The QCD era is conjectured to start from a very high energy scale, i.e, very small $\alpha_s$, and ends up when $Q\rightarrow \Lambda$, i.e, at divergent $\alpha_s$. As a matter of precaution, we assume that the QGP matter exists in a much narrower energy scale, up to just $3-5$ times $\Lambda$. According to the recent lattice QCD simulations \cite{karsch07}, $\Lambda$ can be localized at $T_c\simeq0.2\,$GeV, which sets the {\it latest} end of the QCD era. Also, we know that the bulk viscosity is about to diverge at $T_c$, and decays with increasing temperature. Based on these two ingredients, we set some {\it narrow} limits to the validity of including finite bulk viscosity in the QGP era.

\begin{figure}[htb]
\includegraphics[angle=0,width=10.cm]{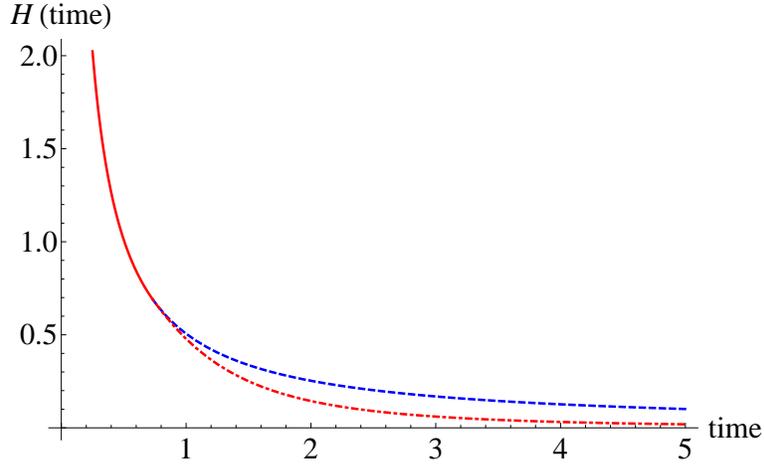}
\caption{\normalsize Numerical solution for the Hubble parameter $H(t)$ for a bulk viscous fluid with
finite bulk viscosity in the full causal Israel-Stewart theory (dashed--dotted bottom curve). The dashed curve (top) represents $H(t)$ in the QCD era with non--viscous background matter. The physical units in both axis are GeV. The solid curve (left) gives the evolution in the pre--QCD eras, where vanishing viscosity characterizes the background matter. }
\label{fig:3a1}
\end{figure}

In the left panel of Fig.~\ref{fig:1a1}, the evolution of the Hubble parameter $H$ is given for different values of the first--order bulk viscosity coefficient $\alpha$. Obviously, the deviation from the non--viscous evolution increases with increasing $\alpha$. The vertical lines determine when or where QGP matter is becoming dominant in the early Universe. Pre-- and post--cosmological eras are probably characterized by EoS's that might be different from the ones used in the present work. Under these assumptions, the viscosity effects are assumed to be localized within the QGP era. In the right panel, the evolution of the time derivative of $H$ is presented for different values of $\alpha$. As shown in Eq.~(\ref{eq:3}), $\dot H$ is the sum of two terms. The first term gives a ratio of acceleration, $\ddot a$, to a scale parameter, $a$. The second term is the square of $H$ itself. The sum of these two terms is negative at small $t$. At large $t$, it the two terms are identical not only qualitatively, but also quantitatively. That the Universe is obviously accelerating, $\ddot a>0$, is also shown by the values of the scale parameter $a$, which increase with increasing $t$. Therefore, we conclude that including finite bulk viscosity seems to affect the expansion of the Universe in a significant way. The evolution of the Hubble parameter likely decays with a fast rate, that increases with increasing $t$. Furthermore, we point out that increasing the values of the bulk viscosity coefficient $\alpha$ accelerates the decay of $H$.

A numerical solution for the evolution equation, Eq.~(\ref{eq:1}), at finite bulk viscosity, is presented in Fig. ~\ref{fig:3a1}. Here, we take into consideration the barotropic dependence of $\xi$, as it has been obtained in the lattice QCD simulations. We do not make any further assumptions on it. Also, the dependence of the relaxation time $\tau$ on the energy density has not been modified. These two ingredients have been strictly implemented. The consistency of these barotropic EoS's with the laws of thermodynamics has been discussed in Ref.~\cite{tawuro}. It seems that all barotropic EoS's used in the present (and also in the previous work \cite{earlyQGP,TawCosmos}) are thermodynamically consistent. Therefore, we numerically solve the evolution equation for the Hubble parameter. The analytic solution has been obtained in Ref. \cite{earlyQGP}. This solution is sensitive to the particular assumptions made in order to solve the second Abel--type non--linear non--homogeneous differential equation. To avoid their effects, we consider the numerical solution to the evolution equation. The boundary conditions required for the numerical methods are defined at the boundaries of QGP era in the early Universe.

A comparison between the time evolution of the Hubble parameter $H$ in the non--viscous and viscous background matter is presented in Fig.~\ref{fig:3a1}. The top dashed curve represents the evolution of the non--viscous cosmological matter, given by Eq.~(\ref{eq:3}). The dotted--dashed curve gives the evolution of the  viscous cosmological matter. In order to compare with the results presented in Fig.~\ref{fig:1a1}, the {\it earliest} (left) time boundary has been zoomed in. By comparing the numerical and analytical results leads us to the conclusion that the two methods (perturbative and numerical) agree in reproducing the evolution of $H$. Also, we conclude that the time evolution of $H$ for viscous cosmological matter is faster than the evolution for non--viscous cosmological matter. Such a difference can be estimated, quantitatively. Therefore, essential cosmological consequences are to be expected from the inclusion of bulk viscous effects in the description of the cosmological expansion of the Universe.

\section*{Acknowledgments}

We would like to thank to the two anonymous referees, whose comments and suggestions helped us to significantly improve the manuscript. TH is partially supported by a grant from the Research Grants Council of the Hong Kong Special Administrative Region, China (Project No. HKU 701808P).

\end{document}